# A Blueprint to Design Curriculum and Pedagogy for Introductory Data Science

Elijah Meyer

Department of Statistics, North Carolina State University

and

Mine Çetinkaya-Rundel

Department of Statistical Science, Duke University

August 5, 2025

### Abstract

As the demand for jobs in data science increases, so does the demand for universities to develop and facilitate modernized data science curricula to train students for these positions. Yet, the development of these courses remains challenging, especially at the introductory level. To help instructors to meet this demand, we present a flexible blueprint that supports the development of a modernized introductory data science curriculum. This blueprint is narrated through the lens and experience in teaching the introductory data science course at Duke University. This is a large course that serves both STEM and non-STEM majors and includes the incorporation and facilitation of technologies such as R, RStudio, Quarto, Git, and GitHub. We identify and provide discussion around common challenges in teaching a modernized introductory data science course, detail a learning model for students to grow their understanding of data science concepts, and provide reproducible materials to help empower teachers to adopt and adapt such curriculum at their universities.

# 1  Introduction

The demand for data science is here. An estimated 11.5 million new data science jobs are projected to be created by 2026, while employment of data scientists is projected to grow by 36 percent from 2021 to 2031 ([U.S. Bureau of Labor Statistics 2022](#)). As job market demands increase, so do the demands in academia. These demands put pressure on universities and data science educators to both offer courses in data science, and to be prepared for data science class sizes to increase ([Redmond 2022](#)). This comes with a commitment to developing modern, nimble, scalable curricula in order to prepare students for this fast-evolving field. Currently, there are associations that provide recommendations on what a modern undergraduate data science course looks like.

The National Academies of Sciences, Engineering, and Medicine offers visions for undergraduate data science programs, courses, and curricula, including: embracing data science as a vital new field; focusing on attracting students with varied backgrounds; preparing to evolve programs over time; have programs be continuously evaluated ([National Academies of Sciences, Engineering, and Medicine 2018](#)). The Curriculum Guidelines for Undergraduate Programs in Data Science provides six major recommendations as to what practitioners of data science should be competent in: computational and statistical thinking, mathematical foundations, model building and assessment, algorithms and software foundation, data curation, communication and responsibility ([De Veaux et al. 2017](#)). Additionally, the Association for Computing Machinery Education Council's Data Science Task Force explores and expands discipline-specific conversations around the field of data science ([Danyluk et al. 2021](#)). This task force acknowledges that data science curricula can be flexible, but suggests that data science curricula should include applications designed towards building skills in computing, statistics, machine learning and mathematics.

Despite current curricular recommendations, academics are still wrestling with what a modern data science curriculum should look like (Schwab-McCoy et al. 2021), and how it can be effectively taught. Further, due to existing implementation challenges such as large class sizes, lack of training, and misaligned learning objectives, many of the aforementioned recommendations are not being implemented, with the majority of current curricula largely focusing on how to model data (Donoho 2017, Schwab-McCoy et al. 2021). The curricula and classroom decisions become even less clear on what context constitutes a well-developed modernized introductory data science course.

While establishing curriculum consensus for introductory data science is hard to reach, providing a blueprint for a functional introductory curriculum and classroom is helpful for the education community. This paper sets out to provide that blueprint and focuses on the creation and implementation of a modernized introductory data science course. We provide reflections and suggestions on how to incorporate, create, implement, and facilitate a modernized data science course using the Data Science in a Box curriculum (Cetinkaya & Ellison 2020). This includes addressing realistic challenges instructors may face, such as having large class sizes, teaching students with little to no data science backgrounds, and incorporating technology (e.g., R, RStudio, GitHub) into such a course. Within this, we provide a framework that outlines how we develop students' understanding of introductory data science concepts and provide reproducible materials to help instructors create, develop, and facilitate their own introductory data science course at scale.

These discussions are through the lens of the data science curriculum for an introductory data science course at Duke University titled Introduction to Data Science and Statistical Reasoning (STA 199). While we have chosen to speak through the lens of STA 199, the goal is not to teach others to create this exact course. Instead, this course is used as a foundation for experiences that we can speak to, and others can learn from to create a data science course tailored to their teaching styles and situations.

This course is designed for large class sizes that enrolls students with little to no statistics, data science, or programming experience. By the end of this course, students are expected and able to tidy, reshape, visualize, model, and communicate with data in a reproducible manner. Detailed learning objectives of this course include learning to explore, visualize, and analyze data in a reproducible and shareable manner with R ([R Core Team 2023](#)), using RStudio as the integrated development environment ([Posit team 2023](#)), reproducible data analysis with Quarto ([Allaire, J.J. and Teague, Charles and Scheidegger, Carlos and Xie, Yihui and Dervieux, Christophe 2022](#)), and Git and GitHub for version control and collaboration ([github 2020](#)). In the following sections, we provide a flexible "first-person" perspective blueprint for course creation, modernization, and implementation through the lens of our introductory data science course at Duke University.

## 2  The course

There are no prerequisites for students to take STA 199. Students who enroll in this course are predominantly undecided on their major and have a variety of interests that span across subjects such as public policy, biology, computer science, nursing, and statistics. Commonly, most students have little to no statistics, data science, or coding experience. In a typical semester, this course seats roughly 150 students, and has seated up close to 300 students in a single section. These numbers are considered to be large by all measures. Students' lack of coding experience and managing large class sizes are identified as two common hurdles by faculty when trying to create and instruct an introductory data science course (Schwab-McCoy et al. 2021, Kokkelenberg et al. 2008).

The design of this course is largely influenced by the *cake first* design principles (Cetinkaya & Ellison 2020). This includes captivating students interests in data science early by showing students the "end result". The end result may include showing students sophisticated and meaningful graphs that they will be able to create by the end of the semester. We then facilitate and encourage students to make their first meaningful data visualization on day one.

This course is built on four large-scale learning objectives: learn to explore, visualize and analyze data in a reproducible and shareable manner; gain experience in data wrangling and munging, exploratory data analysis, predictive modeling, and data visualization; work on problems and case studies inspired by and based on real-world questions and data; learn to effectively communicate results through written assignments and project presentation. These objectives are accomplished through interactive lectures and labs that present content, problems, and case studies inspired by and based on real-world questions and data.

When teaching, instructors are committed to helping students build up a foundation of knowledge that sets the stage for each student to not only accomplish tasks, but train their brain to work through complex statistical and coding problems (see Figure 1).

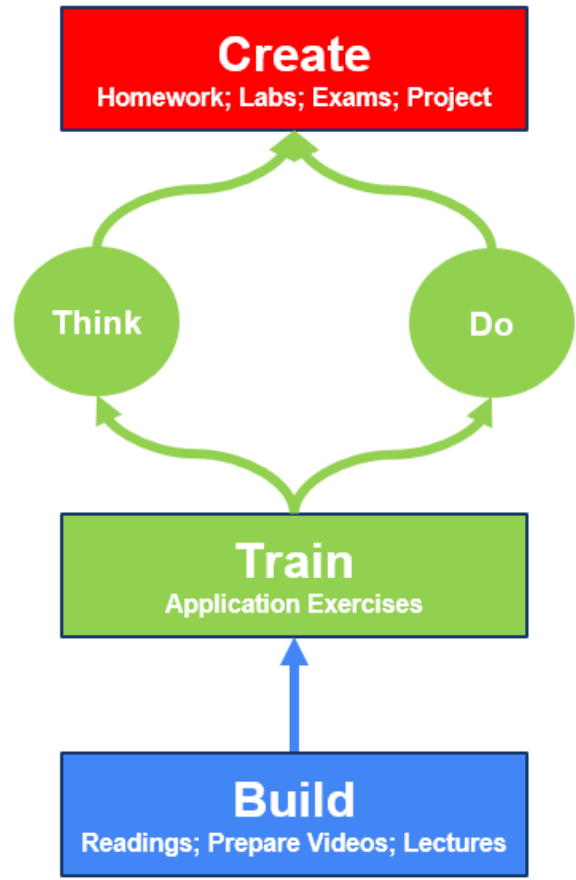

Figure 1: Data Science Student Learning Model for STA 199

This model yields five components that are woven into the student learning structure of this course. In the initial *Build* phase, students are first introduced to content through assigned readings, videos, and lectures. During this phase, students learn through multiple modes of content delivery, and create a foundation of information that they can continue to build upon. We both do and recommend assigning readings from R for Data Science, 2e ([Wickham & Grolemund 2023](#)) and Introduction to Modern Statistics ([Çetinkaya-Rundel & Hardin 2021](#)). The videos, created by an instructor of STA 199, act as an additional resource to compliment the reading and lecture material, and are expected to be viewed prior to class.

Students then shape their existing knowledge by *Training* their brain through interactive hands on in-class activities called application exercises (AEs). During these exercises, students are given in-class opportunities to solve problems individually, as a group, and at the class level. During this phase, there is emphasis placed both on the *Doing* (steps needed to accomplish the task), and the *Thinking* (how to accomplish similar tasks in the future).

This culminates into the *Create* phase where students demonstrate what they have learned, and connections they can make with the material. The material they create are in the form of assessments that includes homework assignments, quizzes, labs, exams, and a project. This model is designed to both situate the student and guide the instructor in facilitating an overall quality learning experience. An example prepare video and AE can be found in Section 6.

This model is integrated with the *Data Science in a Box* curriculum (datascienceinabox.org). Topics taught fall under four major units: Exploring data; Data science ethics; Making rigorous conclusions; and Looking further (Cetinkaya & Ellison 2020). In the first two units, students are introduced to R, RStudio, Quarto, Git, and GitHub. While exploring data, students start to create data visualizations and learn how to both import and manipulate data to be better suited for modeling. Next, multiple classes are spent having conversations around and facilitating activities on the topic of data ethics to help ensure students start developing the skills necessary to be responsible researchers. In the next two units, students extend their investigations and understandings of data into a modeling framework. Specifically, students fit a variety of models (simple linear regression, multiple linear regression, logistic regression), and learn the fundamentals of hypothesis testing and confidence intervals. The looking further section includes independent topics that the instructor can choose to teach, typically at the end of the semester. Topics have included cryptanalysis and genetic forensic analysis, visualizing spatial data, Bayesian inference, creating interactive web applications with Shiny (Chang et al. 2023), text analysis and text modeling. The goal of these lessons are to provide an opportunity to have students learn about topics that interest them in a "no-stakes" environment, continuing to excite and inspire students into a career in statistics and data science.

# 3 Implementation

*"Education without application is just entertainment."—Tim Sanders;*

The content and learning goals of a course are, of course, very important. However, the success of a course depends equally, if not more, on the details of implementation. This includes strategies for constructing a teaching team to help instruct the course as well as technology and pedagogical choices. As we discuss this, we also provide reproducible code that aims to support newer instructors who are developing or teaching an introductory data science course; instructors with courses that are increasing in size; and instructors who want to implement more technical tools into their curricula and classroom found in Section 6.

## 3.1 Teaching team

We structure a teaching team to help account for the difficulties a large class size can bring. Our teaching team consists of one instructor and multiple teaching assistants (TAs). The responsibilities of any TA is to both support the instructor in charge of the class, and support the students in the classroom. These TAs range from undergraduate to PhD level students, and vary in teaching experience.

Once training is complete, TAs are assigned roles that indicate their responsibilities during the semester. These roles include *course organizer*, *head TA*, *lab leader*, and *lab helper*. Often, these roles are given based on the academic level of student, with more academically experienced TAs taking on the roles of course organizer, head TA, and lab leader, where as students with less experience (e.g. undergraduate students) take on the role of lab helper.

Lab sections are held once a week, and are facilitated in person by both a lab leader and lab helper. During lab, students apply the concepts discussed in lecture to various data analysis scenarios, with a focus on computation. The responsibilities of a lab helper are supporting both the students and lab leader as they see fit. Examples of support may include setting up the classroom before class, or conducting small group conversations when students have questions about the material.

The lab leader is responsible for running the lab. This may involve giving a brief introduction and wrap up of lab content, as well as being able to answer questions and facilitate conversations among students about the lab assessment material. In addition, both lab leaders and helpers must hold two hours of office hours each week and have grading responsibilities assigned throughout the semester.

Head TA responsibilities can generally be categorized into the following categories: administrative and pedagogical. Administrative responsibilities include the organization and distribution of TA responsibilities throughout the semester. It is imperative that the head TA and instructor clearly communicate expectations with each other to establish exactly how rules and responsibilities are assigned to TAs. This includes distributing grading assignments and deadlines to both lab helpers and leaders, weekly. The head TA also makes sure that all TAs complete grading within a week and spot check the grading accuracy and quality of all written feedback given. Other administrative duties include reminding other TAs about bi-weekly payroll deadlines and ensure TAs are working their alloted hours per week (and not more). Pedagogically, head TAs are responsible for creating or reviewing answer keys and grading rubrics for homework and lab assessments as the instructor sees fit. Each head TA is also assigned to instruct one lab section during the semester. Before becoming a head TA, there is additional training that specializes head TAs in their administrative responsibilities.

The course organizer is expected to work across each section of STA 199, instead of working with a single instructor. Their responsibilities include creating and/or collaborating with other TAs and the instructor on rubrics for homework and lab assignments. Additionally the course organizer, along with the instructor, answers real time questions virtually during labs asked by lab leaders. Questions often range from content related to technical questions about GitHub and R. Finally, the course organizer is responsible for handling all requested assignment extensions from students. This includes filing away student exemptions, providing extensions for extreme circumstances, and enforcing the late work policy outlined in the syllabus when necessary.

We typically have one course organizer, one head TA, six lab leaders, and six lab helpers per section. Although this is our proposed team structure, we acknowledge different universities have different sets of resources, and want to emphasize that there is great flexibility in coming up with how a teaching team is built and operates for an introductory to data science course. Among any team, we encourage a system designed to alleviate the grading responsibilities of a large class from a single individual, dispersed into many among the team. When grading, it is suggested that each individual is properly trained on how to grade assessments, and expectations on how to provide feedback are clear. Unclear feedback given has often been a point of contention from students in the past. For each assignment, we recommend having one member of the team grade one problem across the entirety of the class roster. This helps adhere to more consistent grading, and encourages more grading questions to be asked as they arise, earlier in the grading process.

Through our experiences, it has been imperative that everyone within the team is communicating with each other. A team with many different roles poses risk for the instructor to be unaware of how or what decisions are being enacted at the grading and lab levels of the course. Thus, it is recommended that the instructor trains everyone on the teaching team to use a communication system that allows every member to communicate any questions they may have, or decisions they make, to the entire team. In the past, we have used the software *Slack*, with appropriately named channels such as *grading-questions*, where TAs can post student answers and questions about grading so the instructor can clearly respond with their expectations. Further, it should be noted that the head TA should not be treated as a "bridge of communication" for the instructor to the rest of the teaching team. It is critical that that the instructor is in consistent contact with all members of the teaching team in making sure all lab leaders and helpers understand the course content, how to grade, and know what's expected of them in their assigned role. We recommend holding a weekly meeting with all members of the teaching team to ensure this. When members are unable to come, it is an expectation that they watch a recorded video of the meeting and reach out if they have any questions about what was discussed.

The utilization of a teaching team helps provide students prompt feedback on assessments, which is critical in developing students understanding. We have decided to implement a teaching team, and share the responsibility of grading to increase turn around time. Having a teaching team also provides students multiple opportunities to receive help on material from a variety of perspectives. It is understood that having a large teaching team may be unrealistic, depending on the university resources. We aim to provide a structure above that can be adapted to size of your teaching team. In the absence of a teaching team, it may be advantageous to implement group assignments even earlier in the semester. This strategy helps mitigate the sheer quantity of grading from a large class size, as well as lay the foundation for students to learn coding together from multiple perspectives.

## 3.2 Technology

We use R, RStudio, and GitHub to create interactive lessons, and assign pre-created assessments to individual or groups of students. In the following sections, we detail the computing infrastructure and share reproducible resources needed to do so. This includes details and examples with R, RStudio, and GitHub, from the instructor's perspective, to set up lectures, AEs, labs and homework. First, we justify our decision to use certain technologies in STA 199 before detailing necessary steps and student information needed so creation can take place.

### 3.2.1 Why GitHub

GitHub is an online hosting version control platform that allows users to manage, track, and store their coding work (github 2020). We choose to use GitHub for STA 199 because it easily and efficiently allows us to create and administer assessments and interactive lessons to a large class size through individual GitHub repositories. Within these repositories, we can provide template code and other resources necessary to best facilitate an interactive lesson where students can code together with both each other and the instructor during a lesson.

This system both calls for and encourages time investment into the creation of assessments and lesson plans. With the time investment needed to create enticing and interactive AEs, it is critical that these resources are easily transferable across semesters. Hosting assessments and AEs on GitHub provides the ability to efficiently manage, update, and re-use materials for subsequent semesters. From the student perspective, exposing them to version control software early in their academic career helps them develop good habits as researchers and emphasize the importance of reproducibility in science.

### 3.2.2 Why R & RStudio

R is a statistical programming language for computing and modeling while RStudio is an integrated development environment for R (RStudio Team 2020). We choose these resources as instructors because they integrate seamlessly with GitHub, providing efficient tools to be used to distribute assessments and in-class activities to students. R is also the coding language STA 199 students learn. There are many benefits to introduce this language from the student perspective. First, they are freely available to download and are currently in high demand in different data science job opportunities. In addition, this programming language has a comprehensive library of packages that allows students to create data visualizations and seamlessly analyze data, aligning with our learning objectives for the course. Further, this software is compatible with online versions that students can access without having to go through a local installment if they do not wish or are unable. There exists online versions that can be set up for students, with capabilities of pre-populating their software with the appropriate packages needed for the semester. Addressing these hurdles for students early is advantageous for any instructor who is working with a large amount of students that are inexperienced with coding.

### 3.2.3 Setting up students: R & RStudio

In an introductory course, it is recommended to minimize student frustration and distraction through the use of pre-packaged computational infrastructure. This is to avoid complications with different operating systems and software versions across students that can potentially dampen their excitement around coding ([Çetinkaya Rundel & Rundel 2018](#)). This also helps alleviate challenges of a large class size, as it is not feasible to expect every student within a large class size to locally install R and RStudio for this course. Alleviating initial student frustrations through these means further provide a more inviting experience to those who have never coded before.

Per this recommendation, students in STA 199 use R and RStudio through a Duke Container. Duke University containers are web browser-based that provides instructors the opportunity to facilitate the use of different software, such as R and RStudio, through an online container instead of needing students to locally download both programs. Additionally, instructors have the ability to manage and install packages students will need for the semester, helping provide a neat and well organized starting experience with a new statistical language.

We want to acknowledge that we provide students instructions on how to download R and RStudio to their local machines, and explain that it is a pedagogical choice to have everyone use the container when first getting started with, and using R in STA 199. The creation and use of these containers is supported by Information Technology (IT) at Duke University. Explicit details of Duke University's RStudio container can be found at https://github.com/mccahill/docker-rstudio.

We also acknowledge that this amount of IT support may or may not be feasible at other universities. There exists other alternatives to help alleviate the initial challenges of R and RStudio set up with students. This may include other virtual machines, the implementation of Posit Cloud, or similar hosted virtual environments like Google Colab, JupyterLab, etc., or creating webR activities that embed interactive R-chunks into the document.

### 3.2.4 Setting up students: GitHub accounts

To participate in AEs and assessments, students must set up a GitHub account. This is done on the first day of class, and often, students are given time during class to sign up. We suggest following tips from "Happy Git with R" ([Bryan & Hester 2020](#)) when creating an account. Some tips include using your real name, and making your name timeless.

Once students create their account, we suggest getting this information from them in a survey. This normally can be done through your learning management system. It is critical to reiterate to students that spelling and capitalization matter when they provide their GitHub username. We suggest asking the question as follows with the following instructions:

```
What is your GitHub username?

Answer this question by ONLY typing your GitHub name and nothing else.
Make sure you triple-check the spelling so I don't add random strangers
to our course organization on GitHub.
```

Once this information is collected, we recommend exporting and storing it as a `.csv` file to be read into R later. We recommend having the following structure of data collected from the survey.

Table 1: Table 1: Example csv for student data

| last_name | first_name | github_username |
|---|---|---|

If you, as the instructor, do not have a GitHub account, you will need to create one as well. Student information will be used to enroll students in your created GitHub organization for your course.

### 3.2.5 Setting up instructors: GitHub organization

GitHub organizations are shared accounts where instructors and students can collaborate across many projects at once. When designing your course, we suggest running it through a GitHub organization that is managed with R and RStudio. This way, this configuration allows for efficient distribution of assessments to individual students or groups of students who are members of your GitHub organization (see Figure 2).

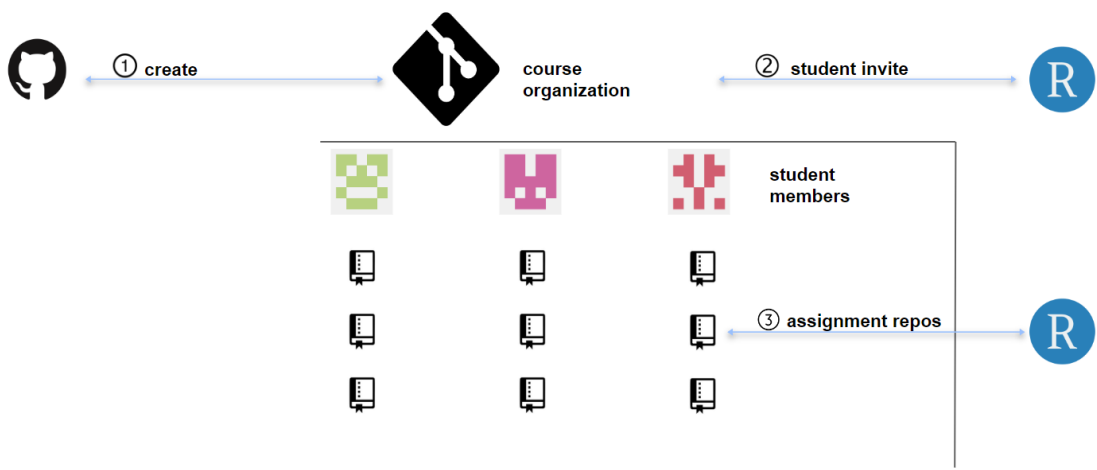

Figure 2: GitHub organization structure for STA 199

### 3.2.6 Create (1)

Using your GitHub account, you can create a new GitHub organization by clicking on your profile icon in the upper right hand corner, clicking *Settings*, *Access*, *New Organization*. It's suggested to name this organization the name of your class and the current semester you teaching in (e.g., STA 199-s23). Once your organization is created, you can use packages within R and RStudio to invite students to enroll.

### 3.2.7 Student invite + Assignment repositories (2 & 3)

Supplemental R code to compliment this section can be found in Section 6. This includes code on how to invite students into your personal GitHub organization, and distributing assessments to students or groups of students via GitHub.

STA 199 is operated through a GitHub organization where students have the capability to receive and clone activities onto their personal computer. To manage and maintain your GitHub organization, we will use a myriad of functions from the `gh_class` package. Students can be invited into your GitHub organization using the `org_invite()` function along with the name of your organization and student GitHub usernames. Once the invites are sent, students will need to accept the invitation to the organization. After students have accepted their invitations, instructors can distribute created assessments to students within the class GitHub organization by using the function `org_create_assignment()`.

Additionally, assessments can be distributed to groups of students acting as a team (See Figure 3).

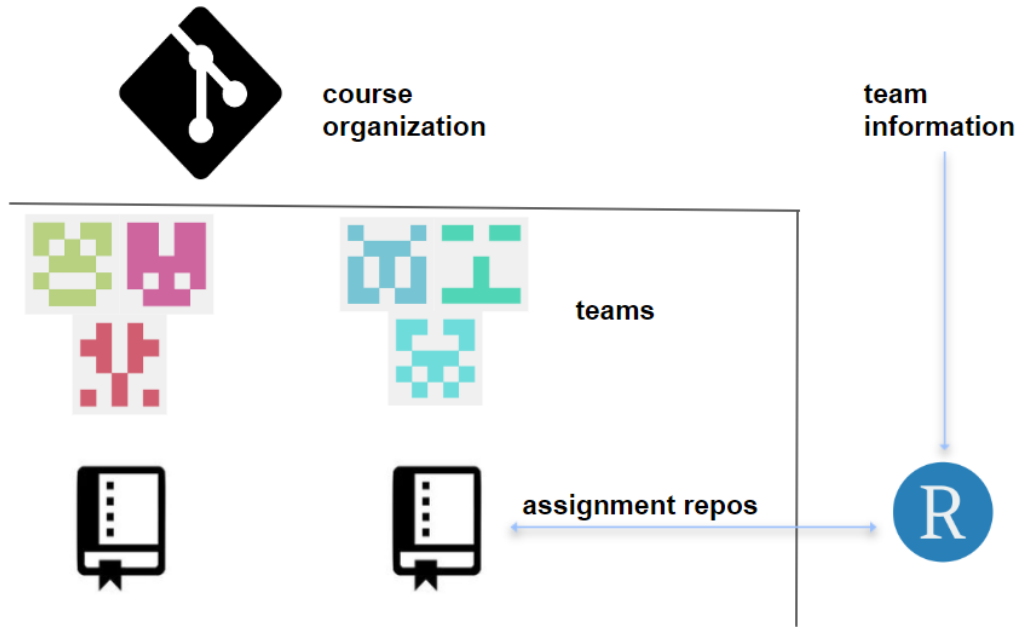

Figure 3: GitHub organization structure for team assignments in STA 199

This is often advantageous for group projects, and to allow students the opportunity to use GitHub as a collaboration tool. To accomplish this, student team information needs to be collected in conjunction with their GitHub username in a format interpretable using R. We add this information to the following roster document.

Table 2: Table 2: Example csv for team data collection

| team_name | last_name | first_name | github_username |
| --- | --- | --- | --- |

We can use these new data and change the code found in `org_create_assignment()` function to create repositories for each team. That is, each individual student will receive a team repository that each member has access to.

Streamlining your course through R, RStudio, and GitHub greatly alleviates common chal-lenges that arise when working with a large class size, such as assessment creation and distribution in both a hands on and reproducible manner. How to create an assignment to distribute is detailed in later sections.

# 4 Pedagogy

*Teaching is both a science and an art.; How and why learning takes place;*

In STA 199, we have chosen a combination of teaching methods and learning assessments to help prepare introductory data science students with the tools they need to be successful outside of university or in future coursework. Our pedagogy includes a combination of lectures, facilitating in-class AEs, and running a lab that focuses more on computation. Figure 4 shows a typical week in the semester in STA 199.

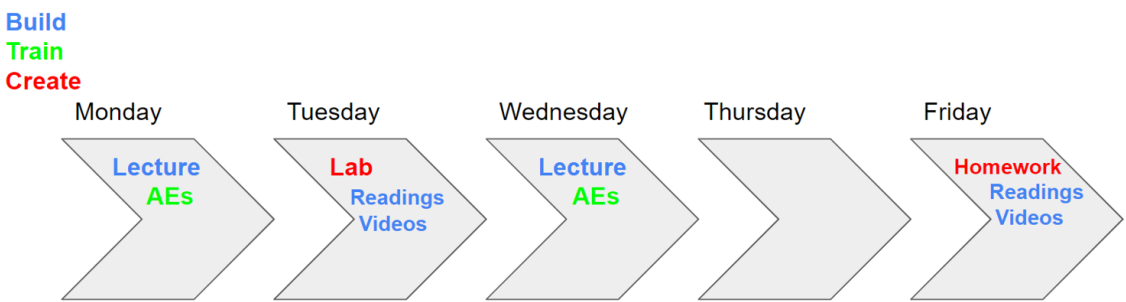

Figure 4: Pedagogical choices during a typical week in STA 199

In the following sections, we detail our justification of, experiences with, and provide suggestions when creating and implementing such pedagogy.

## 4.1 Lecture & AEs

Class is held twice a week during a typical week in the semester. Class is a combination of lecture and interactive AEs facilitated by the instructor. Hands on interactive AEs are the backbone of learning in STA 199. We have found that providing students with the opportunity to code in real-time during class keeps them more engaged and eager to continue learning about the material, versus simply lecturing for a 75-minute period. AEs are designed to be a low stakes assessment where students can earn full credit by completing them in class.

The purpose of these AEs are to give students the opportunity to learn and fortify their understanding of the material by applying the statistical concepts and in real-time. Students are expected to bring their laptops to class to participate in AEs. This expectation needs to be made clear prior to the first day of teaching, and the instructor is encouraged to demonstrate how to clone a GitHub repository that contains the AE.

When first creating an AE, we suggest streamlining the process through a GitHub template repository cloned into RStudio as a project. We choose to use Quarto, within RStudio, to create AEs (Allaire, J.J. and Teague, Charles and Scheidegger, Carlos and Xie, Yihui and Dervieux, Christophe 2022). This choice is intentional, to provide students the opportunity to practice writing code and answering questions in a reproducible format supported by R and RStudio. As highlighted in *Implementation*, this system, streamlined through GitHub, helps the instructor navigate large class sizes, and provides a space to to store and come back to your AEs in subsequent semesters. This space provides a strong environment for students to learn version control, and saves instructors time in the future semesters, as AEs are organized to be re-tooled instead of recreated. This saves instructors valuable time and energy before and during the semester.

When designing questions for an AE, we suggest creating a mix of coding and concept questions (e.g., fill in the code blanks, short response questions) that encourage students to follow along with instructor demonstrations, and also provides students an opportunity to answer questions on their own. During live coding, there are always situations where students can fall behind in class, limiting the amount of value they may receive from the exercise. Thus, we have come up with strategies to try and mitigate this situation.

We suggest that questions are scaffolded (e.g., fill in the blank), especially at the beginning of the semester, so students are not overwhelmed trying to code from scratch in real time. This tends to ease tension for those first learning code, and help instill confidence within students when the code runs.

Additionally, we encourage to give students opportunities to continue following along coding even if they fall behind. For example, if students fall behind or type incorrect code when trying to initially follow along, we try to provide students access to correct code in a separate document (or later in the AE) that grants them the ability to continue engaging for subsequent questions. This is especially important at the beginning of the semester to minimize students' frustration around a new coding language. Moreover, we suggest clearly labeling where students are expected to type out answers in text or code throughout the exercise to further streamline their involvement and continue to minimize frustration. An example of an AE used to help teach tidying data can be found in Section 6.

The amount of time dedicated to lecture and AEs can and should vary by instructor. Typically, we allot the first 15 minutes of a 75 minute class for lecture. During lecture, students are introduced to new content, or content from the prepare videos, that will be re-enforced in the AE. Other lecture strategies include starting class with a warm-up question to get students thinking about today's material and provide a "no-stakes" assessment on the required preparation readings and videos.

We highly recommended designing AEs with built in time to have students work on their own. The amount of time can largely depend on the question being asked, and the teaching style of the instructor. We have found that multiple 3-5 minute blocks for students to answer questions without the instructor's guidance works well. This has been especially effective if the code is then written together as a class, built off student responses, to provide immediate feedback to students before moving on to the next set of material or questions.

To help maximize engagement of such a large class size, we allot more class time to individual and team-based AEs where students are asked to code on their own, together, or along with the instructor. This is followed by a roughly 5-10 minute allotment for a wrap up lecture, summarizing the highlights of AE.

It is highly advised that live coding sessions be the main staple of your classroom when teaching students. Live coding sessions continue to be accepted as one of the best pedagogical practices for teaching coding (Selvaraj et al. 2021), and have been met in our classrooms with overwhelming positivity from students with varying degrees of coding and statistics experience:

- "*The in-class AEs are extremely helpful for understanding the concepts and programming; AEs helped solidify concepts and gave time for practice.*"

- "*I think the AEs's are very helpful and I love that it is very hands-on and easy to follow along.*"

- "*I really enjoy the AE's and that you take the time to walk us through the code and answer questions.*"

- *"I like how the AE gives us a chance to practice the skills on our own after class as well."*

For AEs to be successful, we suggest spending the first couple classes establishing a routine with your students. This routine ensures that students clone the AEs prior to the start of class, and understand that the expectation is to "learn through doing" in live coding sessions. This routine can be built even if not every student has accepted the invitation to join the classroom organization. We suggest making the first couple AEs in a public repository, so each still have access to the material. We additionally suggest making this a template repository to avoid having to introduce the idea of forking.

Typically, we make AEs worth completion points that total to be 5% of a student's end of semester grade. For students that do not show up to class, we make AEs due three days from when the AE was assigned. At the end of the semester, if student's have completed 80% of AEs, they earn a 100% for their grade. Students turn in AEs by pushing their answers up to their GitHub repository. There have been mixed responses from both students and instructors on assigning a grade to AEs. Disadvantages include not currently having an efficient way to implement a hard deadline that students must adhere to, without removing push access from their AE GitHub repository all-together. Further, if an instructor does not finish an AE in class, students who did not attend class can be easily confused on what's expected of them to complete. We suggest tailoring the decision of grading of AEs to your individual course as you see fit.

## 4.2 Labs

Labs for STA 199 meet once a week for 75 minutes, and are facilitated by lab leaders and lab helpers. The purpose of labs are to allow students to apply concepts found in prepare material, lecture, and AEs, to various data analysis scenarios. Lab section sizes are kept to around 30 students so each have more of an opportunity to converse with each other, the lab leader, and lab helper, when working through the lab assessment.

Much like AEs, the instructor or head TA creates and clones a lab GitHub repository for all students that contains lab content. In the lab GitHub repository, we choose to create a starter document with places for students to write code under specific question numbers. The actual lab assessment questions that this repository corresponds to can be found on the class website that is referenced during lab time.

Roughly one-third of the way into the semester, students are assigned to groups to complete a class project. We suggest strategically assigning groups based on a collection of the following information

- Declared or Intended Major

- Year in School

- Suggested times they work on school assignments

From our experience, groups who have significantly different years in school across students have more friction, and tend to work less well together than students that have more similar years in school. Additionally, we highly suggest pairing up students that share common interests using their intended or declared major in school. The class project assigned is an open ended research project where the group collectively decides on a project topic. We have found that students can feel disengaged or left out of the group if they have different interests than the others, and their interests are not reflected when picking a project topic. When groups are assigned, each group during a lab are tasked to come up with a team name. This team name will be the name used to create their team repositories for the remaining labs.

From this point forward in the semester, we choose to have all labs be completed as group work. Introducing group work during labs and through a class project can help students learn from different perspectives, practice their communication skills, and improve their problem solving skills in the context of data science. The new expectation is that, once groups are formed, one lab assessment will be turned in for each group instead of each individual. This further helps lessen the grading responsibilities to those that are assigned it.

After groups are formed, we give a set of recommendations for groups to help promote a successful and positive group dynamic. This includes:

- Establish a clear line of communication with all members

- Share ideas; Let your voice be heard.

- Teach each other.

- Do not approach group work as a bunch of individual assignments.

In our experience, we have observed that requiring GitHub may promote teams working in other more familiar collaborative platforms to avoid GitHub merge conflicts. We try to disincentivize this through multiple avenues. First, we communicate that each team member must have anywhere from one to three meaningful commits to the team repository prior to any project deadline. Secondly, we situate the importance of learning this collaborative skill in a supportive environment with space to ask questions and work through such conflicts. Lastly, we demonstrate that merge conflicts are a part of collaboration process using GitHub, and explain that not all merge conflicts are a "bad thing."

To further ensure positive group dynamic, we initiate two or three peer review surveys. These peer review surveys provide insight into each group's dynamic and may inform the teaching team of issues that may need to be addressed. Questions within each survey range only having instructor only visibility, to being shared with all team members to promote productive conversations. An example question includes: *Estimate the percentage of the total amount of work/effort done by each member, including yourself. Be sure your percentages sum to 100%.* We suggest adding additional questions as deemed necessary to help best understand how everyone is working together within a group. There have been varying degrees of feedback around question responses being released to the group. Some students have suggested that it helps facilitate productive conversations to make the group project experience a more successful one. Others have revealed that they felt as though they could not be honest in their responses

A new instructor of data science, or one with an increasing class size should also think critically if and how they want to implement group work in their classroom. In our experience, merge conflicts, the number of groups, and group formation have been the most difficult aspects of facilitating group work. Students, especially those newer to coding, are extremely hesitant to create merge conflicts. Despite a lab dedicated to the creation and fixing of merge conflicts, students often express frustration and feel as though they are doing something wrong when merge conflicts occur. Some groups choose to collaborate using other forms of technology (e.g., Google documents) before committing and pushing their finished work onto GitHub. We highly advise against this, and try to incentivise students by emphasizing the practical importance of learning how to work through merge conflicts. Secondly, the sheer number of groups created from large class sizes creates an additional time investment for all members of the teaching team. This includes making sure all merge conflicts can be fixed accordingly, and helping facilitate an appropriate working group dynamic among all groups. Finally, there have been mixed strategies on how to form groups that ultimately have yielded similar results. Typically groups are assigned by the instructor using the aforementioned questions above, while others have allowed students to select their owns groups. In our experiences, both strategies have resulted in both positive and fractured group dynamics. We highly suggest that you consider additional measures and adjust group formation as you sit fit for your course.

There are advantages and disadvantages of having lab leaders and helpers facilitate the lab. Advantages includes having students learning through additional perspectives, while also providing a potentially more inviting atmosphere for questions to be asked about the material. However, disadvantages may include inconsistencies from what is shared in lab vs lecture. It is critical that lab leaders and helpers are trained to both understand and explain concepts consistently to what is being taught during lecture and through the AEs. Additionally we recommend setting the expectation that lab leaders and helpers become familiar with the AE content before facilitating the labs so they can refer students back to resource they are familiar with.

# 5  Discussion

As an emerging discipline, there is a large body of literature that tries to pin down exactly what data science is (Provost & Fawcett 2013). The aim of this paper is to help create and modernize this content into curriculum that best supports computationally efficient data science skills such as reproducibility, data manipulation, and data visualization into an introductory course. Although a consensus for what introductory data science courses should look like is extremely difficult, this paper helps establish a start to a consensus while providing a flexible framework for other instructors to create, advance, and facilitate their own introductory data science curriculum. The content, pedagogical choices, and reproducible materials presented help empower the instructor to teach introductory data science with technology to large class sizes of students with little to no coding experience.

The content structure of this paper assumes you currently, or will be using both R and RStudio when teaching. We acknowledge that there are other prominent coding languages that are used in data science courses. For example, other coding languages, such as Python, are compatible with GitHub and have the functionality to perform basic data wrangling, data visualization, and modeling techniques, and are compatible with online virtual machines. However, a potential drawback could be navigating large classes through assessment distribution. Currently, there is no known functionality in Python that mirrors the flexibility and utility of the `gh_class` package in R for course facilitation.

Four years later, the trickle down effect still holds true when implementing a modernized introductory course (Cetinkaya & Ellison 2020). Students leave this course with an understanding of new statistical concepts and advanced uses of new technology (e.g., RStudio, GitHub, Quarto). This helps better prepare students for future coursework, and gives them more time in their academic career to master such skill sets before entering the workforce. Implementing a course like this also is favorable for the instructor. The technological skills introduced in STA 199 are ingrained into the data science pipeline. Thus, teaching a modernized introductory course helps better prepare instructors to teach other data science courses. Further, this opens up the opportunity for program development. The introduction of advanced technology in the introductory course allows for the upscale of subsequent courses. This includes integrating these technologies into other courses where they were not used, and using the time they were initially introduced for other, more content related purposes.

One of the biggest challenges is creating a course like this is the incorporation of technology. Instructors need to invest time on how to both harness the utility of streamlining a course through, and incorporating the power of technology into their curriculum. We hope the blueprint provided accompanied with supplementary materials aid into this investment and provides a spring board to invest the time and energy necessary to help equip the next set of students with the tools necessary to succeed in an ever-evolving workforce. We believe that a course like this will ultimately be the first domino in creating a strong data science pipeline, grow the field, and equip instructors with the tools necessary to continue bettering the educational experience for themselves and their students.

# 6 Supplementary materials

Supplemental materials for the article includes code to set up your github organization and code to distribute assessments to individual and groups of students. We also include an example video on tidying data with a corresponding AE (with solutions). These materials can be found at https://github.com/ElijahMeyer3/ds-blueprint